\documentclass[times,twocolumn,final]{elsarticle}

\usepackage{jasr}
\usepackage{framed,multirow}

\usepackage{amssymb}
\usepackage{latexsym}

\usepackage[switch]{lineno}

\usepackage{url}
\usepackage{xcolor}
\definecolor{newcolor}{rgb}{.8,.349,.1}
\usepackage[normalem]{ulem}

\usepackage[citebordercolor=white]{hyperref}

\usepackage{amsmath}		
\usepackage{comment}
\usepackage[normalem]{ulem}

\newif\ifhighlight
\highlightfalse

\newcommand{\change}[1]{\ifhighlight \textcolor{blue}{#1}\else #1\fi}

\journal{Advances in Space Research}

\begin{document}


\verso{Korolkov, Baliukin, \& Opher}

\begin{frontmatter}

\title{The Effect of Magnetic Field Dissipation in the Inner Heliosheath: Reconciling Global Heliosphere Model and Voyager Data}

\author[1]{Sergey D. \snm{Korolkov}\corref{cor1}}
\cortext[cor1]{Corresponding author at: Space Research Institute of the Russian Academy of Sciences, Profsoyuznaya Str. 84/32, Moscow 117997, Russia.\\{\it Email address:} sergey.korolkov@cosmos.ru (S.D. Korolkov).}

\author[1,2]{Igor I. \snm{Baliukin}}

\author[3]{Merav \snm{Opher}}

\affiliation[1]{organization={Space Research Institute of the Russian Academy of Sciences},
                addressline={Profsoyuznaya Str. 84/32},
                city={Moscow},
                postcode={117997},
                country={Russia}}
                
\affiliation[2]{organization={HSE University},
                addressline={20 Myasnitskaya Ulitsa},
                city={Moscow},
                postcode={101000},
                country={Russia}}
                
\affiliation[3]{organization={Boston University},
                city={Boston},
                postcode={MA 02215},
                country={USA}}

\received{--- 2025}
\finalform{---}
\accepted{---}
\availableonline{---}

\begin{abstract}

Global ideal magnetohydrodynamic models of the heliosphere typically predict a greatly exaggerated magnetic field pile-up in the inner heliosheath (IHS), the region between the termination shock and heliopause. However, Voyager 1 and 2 observations show only a gradual increase throughout this region. This mismatch is largely attributed to the simplified assumption of a unipolar solar magnetic field in many global models, which neglects the complex, folded structure of the heliospheric current sheet (HCS). The IHS, especially at low heliolatitudes, contains these compressed sector boundaries, widely considered prime locations for magnetic dissipation via reconnection. 
To align global model simulations with observations without incurring the prohibitive computational cost of resolving the kinetic-scale current sheet, this work introduces a phenomenological term into the magnetic field induction equation. This term captures the macroscopic effect of magnetic energy dissipation due to unresolved HCS dynamics. It is designed to mitigate the artificial magnetic pile-up, preserve the topological integrity of the magnetic field lines, and avoid explicit magnetic diffusion. 
This study demonstrates that incorporating a phenomenological dissipation term into global heliospheric models helps to resolve the longstanding discrepancy between simulated and observed magnetic field profiles in the IHS. The proposed mechanism reduces exaggerated magnetic energy (converts it into thermal energy), aligns model output with Voyager measurements of both magnetic field and proton density, and produces the outward shift in termination shock position and a reduction of the IHS thickness. We found that the characteristic time for magnetic field dissipation of about 6 years provides \change{improved agreement with Voyager data in the IHS}.


\end{abstract}


\end{frontmatter}


\section{Introduction} \label{sec:intro}

The region where the solar wind (SW) interacts with the local interstellar medium (LISM) is defined by several key discontinuities. These include the termination shock (TS), where the supersonic solar wind abruptly decelerates, and the heliopause (HP), a tangential discontinuity that separates solar and interstellar plasmas \citep{baranov70}. Under certain LISM conditions, a third discontinuity, the bow shock, may exist in the interstellar medium upstream of the HP \citep[see discussions in, e.g.][]{izmod2009, mccomas2012a, zank2013, zieger2013}. \change{The inner heliosheath (IHS) is the region of compressed and heated solar wind plasma that occupies the volume between the TS and HP.} The most significant insights into this remote plasma environment have been obtained from direct, in situ measurements by the Voyager 1 and 2 spacecraft (V1 and V2), which have provided essential data on the plasma flow, magnetic fields, and energetic particles.

There is a significant discrepancy between these empirical constraints and the predictions of global numerical models. While the Voyager data reveal only a gradual increase in the magnetic field magnitude throughout the IHS, conventional ideal magnetohydrodynamic (MHD) models, which assume a perfectly frozen-in magnetic field, predict a drastic pile-up of the heliospheric magnetic field as it approaches the HP \citep{izmod2015, izmod2020, Michael2018, Pogorelov2017, Pogorelov2021}. This increase in the magnetic field leads to the formation of a magnetically dominated layer ahead of the HP, characterized by a low plasma beta ($\beta < 1$, thermal to magnetic pressure ratio) and a sharp drop in plasma number density, which contradicts the observations. This problem is pronounced in global models that employ a simplified, unipolar magnetic field configuration \citep[e.g.][]{izmod2015, opher2015}. 

The use of this unipolar approach is a common practice that allows for avoiding the computational intractability of resolving the finely structured heliospheric current sheet (HCS), the warped layer that separates the regions of opposite magnetic polarity \citep{Czechowski2010, borovikov2011, opher2011}. This method helps to eliminate numerical diffusion of the magnetic field. However, simulations using this approach do not account for the physical magnetic energy dissipation driven by reconnection events at the HCS. Within the compressed folds of the heliospheric current sheet in the IHS, magnetic reconnection becomes highly active, converting accumulated magnetic energy into plasma thermal energy and thereby regulating the field strength \citep{lazarian2009, drake2010, Drake2017}. 

\change{Attempts to incorporate the HCS in global models using a dipolar field do not fully resolve the issue. In such models, the HCS, being under-resolved, becomes a site of severe numerical magnetic dissipation, that is distinct from a physical one and scales with grid resolution \citep[e.g.,][]{opher2011, pogorelov2025}. This grid-dependent dissipation can eliminate the pile-up but often leads to an overly weak magnetic field near the HP and an entirely thermally dominated inner heliosheath, in contradiction with Voyager observations \citep[see, e.g, Figure 10 in][]{Michael2018}. Furthermore, these numerical effects are exacerbated when more realistic features, such as the tilt of the solar magnetic axis or the solar cycle, are included.}

\change{Therefore, the fundamental challenge extends beyond a specific model type: purely unipolar models suppress dissipation entirely, leading to excessive field strength, while dipolar models with an under-resolved HCS introduce uncontrolled, non-physical dissipation, often leading to insufficient field strength. Neither approach yields results fully consistent with Voyager in situ data. This persistent mismatch underscores the need for a controlled, physics-inspired representation of HCS-mediated dissipation.} 



Explicit modeling of collisionless reconnection in large-scale simulations is infeasible because the physical scale of this process, typically at or below ion inertial lengths, is vastly smaller than the astronomical unit (au) scales characterizing the global heliosphere. The core conflict is the need to resolve microscopic kinetic physics, as captured in Particle-in-cell (PIC) or Vlasov simulations, while also modeling large macroscopic global domains using global MHD codes. This creates a scale-separation challenge in plasma physics that is very difficult to overcome \citep[see, e.g.][]{Shay2025}. 

In this work, we propose a novel approach to incorporate the macroscopic effects of this dissipation within a global MHD framework, without resorting to kinetic-scale resolution. We introduce a simple phenomenological term into the magnetic induction equation, parameterized by a characteristic dissipation timescale. Crucially, the removal of magnetic energy through this term results in a heat source in the plasma's internal energy equation, effectively representing the macroscopic equivalent of Joule heating in collisional plasmas. This approach is designed to systematically reduce the magnetic field magnitude, mimicking the net effect of reconnection within the unresolved HCS folds, without altering the large-scale field topology or introducing spurious numerical diffusion. By calibrating this timescale to achieve agreement with the Voyager observations, we obtain a physically realistic global model of the heliosphere, which bridges the gap between theory and critical in situ observations.

The description of the model used in this study is presented in Section \ref{sec:model}. Section \ref{sec:results} provides the results of numerical simulations, comparison with Voyager data, and discussion. The conclusions of this study are summarized in Section \ref{sec:conclusions}.

\section{Model} \label{sec:model}

For this study, we employed a stationary 3D single-fluid model, in which all charged particles — thermal protons, pickup protons, electrons, and helium ions (He$^{++}$ in the solar wind and He$^{+}$ in the LISM) — are treated as a single conducting fluid governed by the equations of ideal MHD. To retain model simplicity, the neutral hydrogen component is represented by four distinct fluids. Each population originates in a different, thermodynamically unique region of the heliosphere (supersonic solar wind, inner heliosheath, outer heliosheath, undisturbed LISM) and is described by its own set of Euler equations. This approach is consistent with many other global models of the heliosphere \citep[see, e.g.][]{Pogorelov2006}. 

The system for plasma mixture is written as follows:
\begin{align}
&\frac{\partial \rho}{\partial t} + \nabla \cdot (\rho \mathbf{V}) = Q^\rho, \\
&\frac{\partial (\rho \mathbf{V})}{\partial t} + \nabla \cdot \left[ \rho \mathbf{V} \mathbf{V} + \left(p + \frac{B^2}{8\pi}\right) \mathbf{I} - \dfrac{\mathbf{B} \mathbf{B}}{4 \pi} \right] = \mathbf{Q}^m,\\
& \frac{\partial E}{\partial t} + \nabla \cdot \left[ \left(E + p + \frac{B^2}{8\pi}\right) \mathbf{V} - \dfrac{(\mathbf{V} \cdot \mathbf{B})}{4 \pi} \mathbf{B} \right] = Q^E,\\
&\frac{\partial \mathbf{B}}{\partial t} + \nabla \cdot (\mathbf{V} \mathbf{B} - \mathbf{B} \mathbf{V}) = \mathbf{Q}^B, \ \ \nabla\cdot \mathbf{B} = 0. \label{eq:induction}
\end{align}
Here, $\rho$, $\mathbf{V}$, and $p$ are the density, bulk velocity, and pressure of the plasma mixture, respectively, $\mathbf{B}$ is the magnetic field vector, $\mathbf{I}$ is the unity tensor, $E = \rho V^2/2 + \varepsilon + B^2/(8 \pi)$ is the total energy, $\varepsilon = p/(\gamma - 1)$ is the internal energy of the plasma mixture, and $\gamma = 5/3$. 

The system for four neutral fluids is written as follows:
\begin{align}
&\frac{\partial \rho_\alpha}{\partial t} + \nabla \cdot (\rho_\alpha \mathbf{V}_\alpha) = Q^\rho_\alpha, \\
&\frac{\partial (\rho_\alpha \mathbf{V}_\alpha)}{\partial t} + \nabla \cdot \left[ \rho \mathbf{V}_\alpha \mathbf{V}_\alpha + p_\alpha  \mathbf{I} \right] = \mathbf{Q}^m_\alpha,\\
& \frac{\partial E_\alpha}{\partial t} + \nabla \cdot \left[ \left(E_\alpha + p_\alpha\right) \mathbf{V}_\alpha\right] = Q^E_\alpha.
\end{align}
Here, $\rho_\alpha$, $\mathbf{V}_\alpha$, $p_\alpha$, $E_\alpha =\rho_\alpha V_\alpha^2/2 + p_\alpha/(\gamma - 1)$ are the densities, bulk velocities, pressures, and total energies of the H populations ($\alpha = 1,\ 2,\ 3,\ 4$, depending on the region of their origin \change{-- the supersonic solar wind, IHS, outer heliosheath, or LISM). Our choice of the four-fluid description follows the methodology detailed in \citet{Alexashov2005}, which provides a comprehensive analysis and comparison of multi-fluid (1-, 2-, 3-, and 4-fluid) approximations against a kinetic model for neutrals within a purely gasdynamic (non-magnetized) heliosphere framework.}

\change{The system of equations for the neutral fluids (5–7) is solved self-consistently with the plasma equations (1–4).} The source terms ($Q^\rho$, $\mathbf{Q}^m$, $Q^E$, $Q^\rho_\alpha$, $\mathbf{Q}^m_\alpha$, $Q^E_\alpha$) in the right-hand side of the density, momentum, and energy equations (1-3, 5-7) describe the influence of the charge exchange process between protons and H atoms, photoionization, and electron impact ionization. For simplicity, the latter two processes are not considered in the model (therefore, $Q^\rho = 0$). The expressions for the source terms are taken from \citet{Bera2023}. In this work, the charge exchange cross-section from \citet{lindsay2005} was used.


The source term $\mathbf{Q}^B$ is responsible for the effect of magnetic field dissipation:
\begin{align}
\mathbf{Q}^B = 
\begin{cases}
    -\dfrac{\mathbf{B}}{\tau} & \mathrm{\ in\ the\ inner\ heliosheath,}\\
    0 & \mathrm{\ otherwise,}
\end{cases}
\end{align}
where the parameter $\tau$ represents the characteristic timescale for magnetic field dissipation. \change{This term mimics a uniform damping process, and, if acting alone, it would cause the magnetic field to decay exponentially, $|\mathbf{B}| \propto e^{-t/\tau}$, analogous to frictional damping in mechanical systems. So,} including this term reduces the magnetic field strength in the IHS. Physically, dissipation is driven by magnetic reconnection in the compressed folds of the heliospheric current sheet within the IHS \citep{lazarian2009, drake2010, opher2011}. However, other dissipation mechanisms may contribute. Proposed alternatives include solar wind turbulence and kinetic/MHD instabilities \citep{Pogorelov2017}. 

In this work, we introduce an effective characteristic time $\tau$, defined as the magnetic field dissipation timescale, which, for simplicity, is assumed to be constant throughout the IHS. Specifically, the limit $\tau \to \infty$ represents the case of an unmodified (no dissipation) heliospheric magnetic field (HMF). In contrast, the limit $\tau \to 0$, if the $\mathbf{Q}^B$ term is applied in the supersonic solar wind region as well, corresponds to the case of complete, instantaneous magnetic field dissipation, which is analogous to having no HMF. 

The justification for the $\mathbf{Q}^B$ term within the scope of this study is the following. First, we aimed to preserve the magnetic topology, that is, to reduce the field magnitude while maintaining its direction. To achieve this, $\mathbf{Q}^B$ must be proportional to $\mathbf{b} = \mathbf{B}/B$, the unit vector along the magnetic field direction. Second, this term must satisfy the solenoidal condition $\nabla\cdot \mathbf{B} = 0$. The most natural solution that meets these requirements is a dissipation term proportional to $\mathbf{B}$ itself. Additional physical justification for this choice of $\mathbf{Q}^B$ in case of the simplified HCS topology is provided in \ref{app:Justification}.

\change{The phenomenological term $\mathbf{Q}^B$ is designed to capture the macroscopic energy consequence of magnetic reconnection -- the conversion of magnetic energy into thermal energy -- while intentionally forgoing a description of the accompanying microscopic topological change. This is a necessary and practical compromise for a global model, as it provides a simple, controlled mechanism to regulate the magnetic field strength without requiring kinetic-scale resolution of the HCS. A true reconnection process would alter magnetic connectivity and field-line topology, effects that our term does not reproduce. Consequently, the model does not generate the spectrum of kinetic-scale structures (e.g., magnetic islands) expected from reconnection. Incorporating a self-consistent description of such global topological evolution remains a significant and intriguing challenge for future models. Nevertheless, this formulation serves as a stable, minimal, and tunable representation of the dominant large-scale dissipative effect within a global MHD framework.}

The system of equations (1–7) is solved numerically using the operator splitting approach, also known as the method of fractional steps \citep{yanenko1967}. The solution of the full system on the next time step is obtained by two (fractional) substeps:

\begin{enumerate}
    
    \item In the first substep, the system is solved without the dissipation of the magnetic field taken into account (i.e., assuming $\mathbf{Q}^B = \mathbf{0}$). The system of ideal MHD (for plasma mixture) and Euler (for fluid neutrals) equations is solved using the finite-volume high-order Godunov's type scheme that includes 3D adaptive moving grid with discontinuities capturing and fitting capabilities, Harten–Lax–van Leer discontinuity (HLLD) MHD Riemann solver, and Chakravarthy–Osher TVD procedure \citep[analogous to the work of][]{izmod2015}.

    \item In the second substep, we solve the corrected induction and thermal conduction equations in the IHS:
    \begin{align}
    &\left\{\begin{aligned}
    &\frac{\partial \mathbf{B}}{\partial t} = \mathbf{Q}^B = -\dfrac{\mathbf{B}}{\tau},\\
    &\frac{\partial \varepsilon}{\partial t} = -\dfrac{\partial (B^2/(8 \pi))}{\partial t} = -\dfrac{\mathbf{B} \cdot \mathbf{Q}^B}{4 \pi} = \dfrac{B^2}{4 \pi\tau}.
    \end{aligned}\right.
    \end{align}
    This system of equations ensures that the dissipated magnetic energy is converted into plasma internal energy, while conserving their sum $\varepsilon + B^2/(8 \pi)$. It provides a heating mechanism analogous to Joule heating in collisional plasma. Such energy redistribution decreases the total pressure $p + B^2/(8\pi) = [p/(\gamma-1) + B^2/(8\pi)] + (\gamma-2) p/(\gamma-1)$, since the term in the square brackets is conserved and the second term is negative. As a result of the second substep, a new distribution of the magnetic field and thermal pressure is obtained.

\end{enumerate}

All the equations above are written in a general time-dependent form (with time derivatives preserved). However, in this work, we present only the results of the stationary model simulations. The steady-state solution is obtained using the time-relaxation method. The stationary inner (at 1 au from the Sun) and outer (in the undisturbed LISM) boundary conditions used in the modelling are described in \ref{app:boundary_conditions}.

We note that our study employs a multi-fluid treatment of H atoms, in contrast to the kinetic approach used in previous studies of our group \citep[see][]{izmod2015, izmod2020, izmod2023}. To validate this method, in \ref{app:compare} we compare the results of our non-dissipative model ($\mathbf{Q}^B = \mathbf{0}$) with the \citet{izmod2020} model under the same boundary conditions. We demonstrated that the multi-fluid approach is suitable for the present investigation. We find this comparison valuable, as it contributes to the ongoing discussion regarding the applicability of multi-fluid hydrogen models in heliospheric studies.

\section{Results and discussion} \label{sec:results}

The value of the magnetic field dissipation timescale $\tau$ could be derived using, for example, PIC simulations that resolve the microstructure of current sheets with a subsequent averaging of the effect on macroscales. In this work, however, we adopt an alternative, phenomenological approach: we determine the unknown dissipation timescale $\tau$ by calibrating our model results against Voyager spacecraft observations. We consider this approach a crucial first step toward accounting for magnetic energy dissipation in global models, since it provides a valuable estimate for the required dissipation timescale in the IHS.

In the following Section \ref{sec:comparison}, we compare the model plasma distributions obtained for the case $\tau = 6$ years with the Voyager data, which was taken from the NASA Space Physics Data Facility (SPDF) repository \url{https://cdaweb.gsfc.nasa.gov/}. As will be shown below, this value of the magnetic field dissipation timescale provides a reasonable agreement with observations in the IHS. The parametric analysis of magnetic energy dissipation effect on the plasma distribution in the IHS (for different values of the timescale $\tau$) and additional reasoning for the chosen value $\tau$ are presented in Section \ref{sec:parametric}.

\subsection{Comparison with the Voyager data}\label{sec:comparison}

\begin{figure}
\includegraphics[width=\columnwidth]{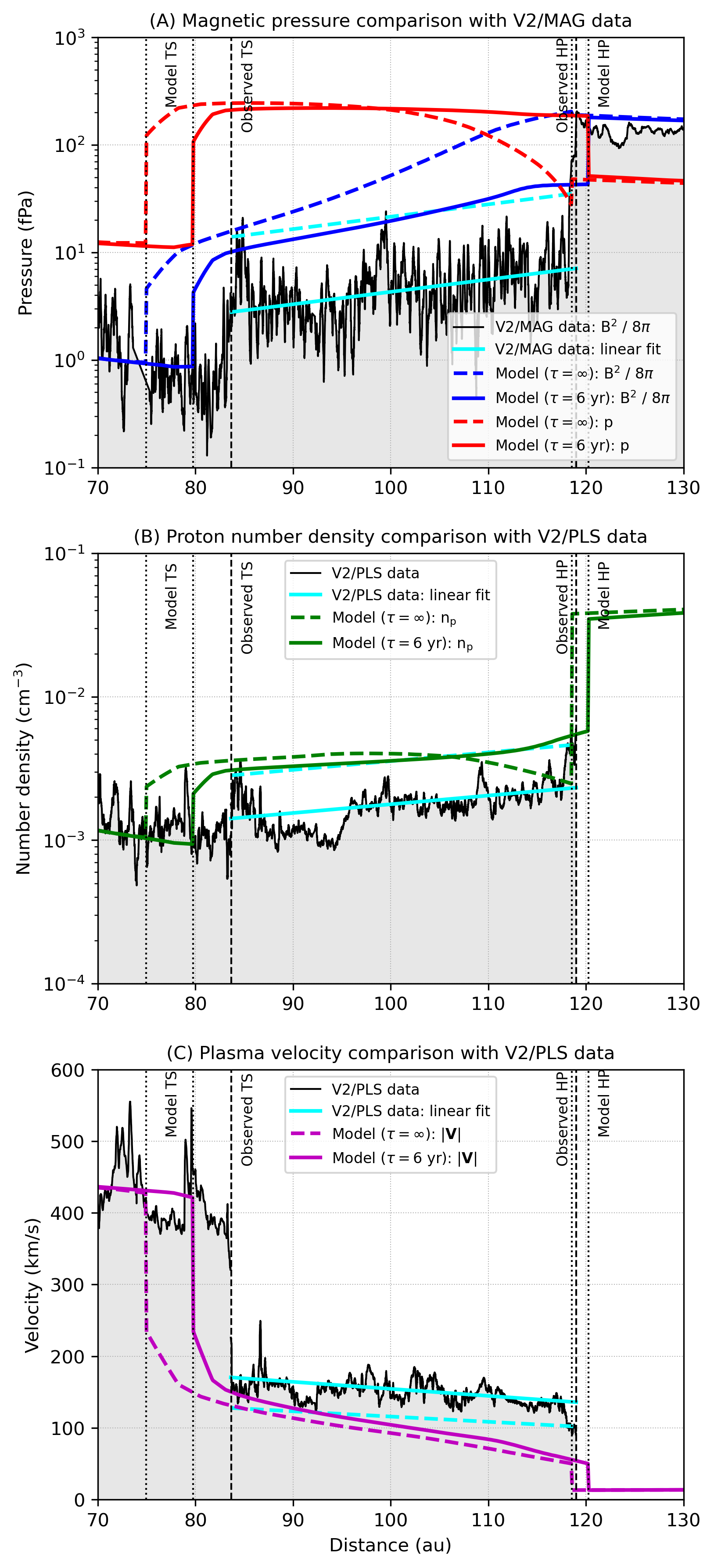}
\caption{
Panel (A) shows the magnetic (blue lines) and thermal (red lines) pressure profiles, panel (B) -- proton number density (green lines), and panel (C) -- plasma velocity module (magenta lines) along the V2 trajectory. The colored lines show the distribution of the model parameters with ($\tau$ = 6 years, solid lines) and without ($\tau = \infty$, dashed lines) magnetic dissipation taken into account. The black lines represent 14-day moving averages of V2 MAG (panel A) and PLS (panels B and C) instrument observations. The solid cyan lines represent linear fits to the data in the IHS\change{, while the dashed cyan lines are the same approximations shifted to the model's values to highlight the consistency of the trends.} The vertical dotted lines show the model TS and HP distances, while the vertical dashed lines correspond to the actual V2 crossings of the TS and HP.
}
\label{fig:v2_comparison}
\end{figure}

Figure \ref{fig:v2_comparison} presents the comparison of the V2 magnetometer (MAG) and plasma spectrometer (PLS) observations with model plasma distributions along the Voyager 2 trajectory. The data values shown in black were averaged over 14-day intervals for better visual representation. The model results are presented with blue, red, green, and magenta lines for two scenarios: the model with magnetic dissipation ($\tau$ = 6 years, solid lines) and the model without dissipation (dashed lines), which corresponds to the case $\tau = \infty$. We note that a direct comparison of the absolute model values with the data is challenging because the model provides (a) only the stationary solution, while observations are essentially time-dependent, and (b) a different width of the IHS (this problem will be discussed below in this section). To facilitate a comparison of trends within the IHS, linear fits to the observational data are plotted in cyan. 

The panel (A) of Figure \ref{fig:v2_comparison} shows the model profiles of magnetic pressure $p_{\rm mag} = B^2 / (8\pi)$ (blue lines) and total thermal pressure $p$ of the plasma mixture (red lines). The figure indicates that in the model without dissipation, the magnetic field accumulates towards the HP. At the same time, the thermal pressure decreases, resulting in the plasma beta $\beta = p / p_{\rm mag} \sim 0.1$. The model with dissipation damps this accumulation, providing an increase in $\beta$ up to $\sim$4.5 at the HP. The close agreement between the model with magnetic dissipation included (solid blue line) and the scaled fit (dashed cyan line) demonstrates that the model accurately reproduces the observed trend.

Figure \ref{fig:v2_comparison}(B) presents a comparison of proton number density distributions: V2 PLS observations (black line), the non-dissipative model (green dashed line), and the dissipative model (green solid line). In the non-dissipative scenario, magnetic pile-up at the HP leads to magnetic expulsion of plasma, resulting in a depletion of proton density towards the HP. This feature is not observed in the data. The comparison with the cyan trend line demonstrates that the dissipative model closely reproduces V2 \change{trends} and effectively eliminates the magnetic expulsion of plasma.

Figure \ref{fig:v2_comparison}(C) shows the profiles of the plasma velocity module from the V2 measurements and the two models. Notably, the velocity profiles in both models are virtually identical, confirming that the considered dissipative process does not directly influence the plasma bulk flow. However, both models show a significant discrepancy with the V2 measurements. The existing difference between the predictions of heliospheric models on the IHS bulk flow and the Voyager data is a known issue \citep[see, e.g.][]{provornikova2014}.

\begin{figure}
\includegraphics[width=\columnwidth]{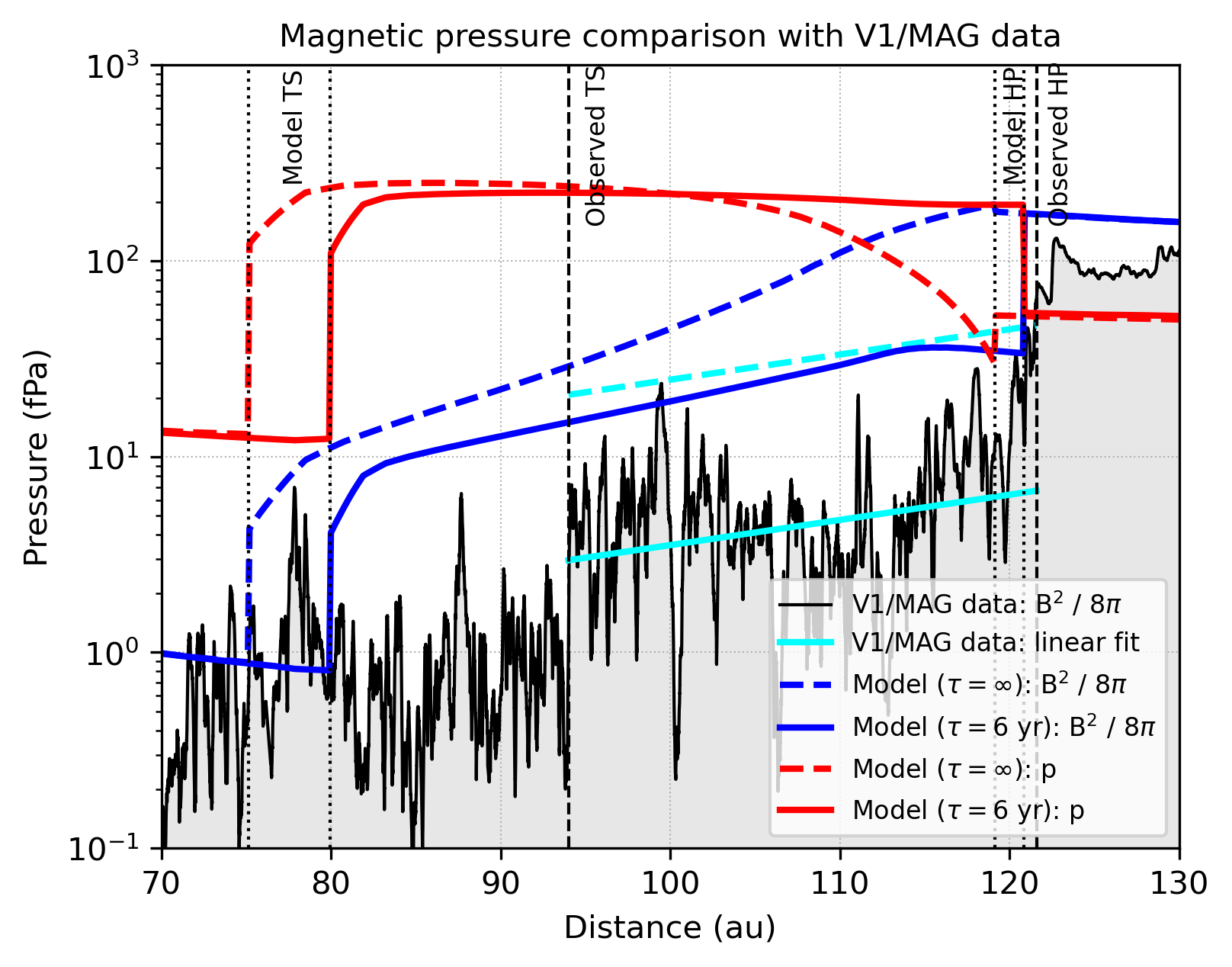}
\caption{
The description is the same as for Panel A in Figure \ref{fig:v1_comparison}, but comparison with the MAG instrument data along the V1 trajectory is shown.
}
\label{fig:v1_comparison}
\end{figure}

Figure \ref{fig:v1_comparison} shows the same comparison of magnetic pressures as in panel (A) of Figure\ref{fig:v2_comparison}, but for the V1 trajectory. The effect of dissipation in the model is similar for the V2 case: the model with magnetic dissipation considered (solid blue line) provides a better match to the observed trend (dashed cyan line) than the non-dissipative model (dashed blue line). However, for the V1 trajectory, the discrepancy in the TS location between the models and the data is even more pronounced than for the V2 case. This mismatch reduces the agreement between the absolute model and the observed data values.

We note that the models produce a wider IHS than observed (in both V2 and V1 directions), which is a well-known problem \citep[see, e.g.][]{izmod2023}. Furthermore, the significant difference in the TS crossing distances between the V2 and V1 directions (84 au vs. 94 au, respectively) cannot be explained by the model with the boundary conditions in the LISM used. In particular, in this work, we utilize the following interstellar magnetic field configuration: B$_{\rm LISM}$ = 3.75 $\mu$G and BV-angle = 60$^\circ$. A different interstellar magnetic field configuration similar to that proposed in \citet{izmod2015} (B$_{\rm LISM}$ = 4.4 $\mu$G and BV-angle = 20$^\circ$) would result in a stronger heliosphere asymmetry and improve the model agreement with the TS locations. However, usage of such a configuration would provide a poor fit to the observed HP distances. 

Another important aspect of this study is the use of a stationary model. The time-dependent models predict significant TS and HP fluctuations \citep[by $\sim$10-15 au from minimal to maximal distances for V1 and V2 directions; see, e.g.][]{izmod2020}. Therefore, considering the time-dependent effects could also help to achieve a better agreement between the model and observed TS/HP locations.

Although this study demonstrates that the magnetic dissipation cannot fully account for the observed IHS width, the dissipative model yields a narrower IHS compared to the non-dissipative case. For Voyager directions, the effect of magnetic field dissipation leads to shifts in the TS positions by $\sim$5 au from the Sun and a reduction of the IHS thickness by $\sim$3 au (for $\tau = 6$ years). As noted in the following section, this change in IHS width is a three-dimensional effect. The narrowing is more pronounced in the V2 direction, less significant in the V1 direction, and nearly absent in the upwind direction.

\subsection{Dependence of the plasma parameters on the magnetic dissipation timescale} \label{sec:parametric}

\begin{figure*}
\includegraphics[width=\textwidth]{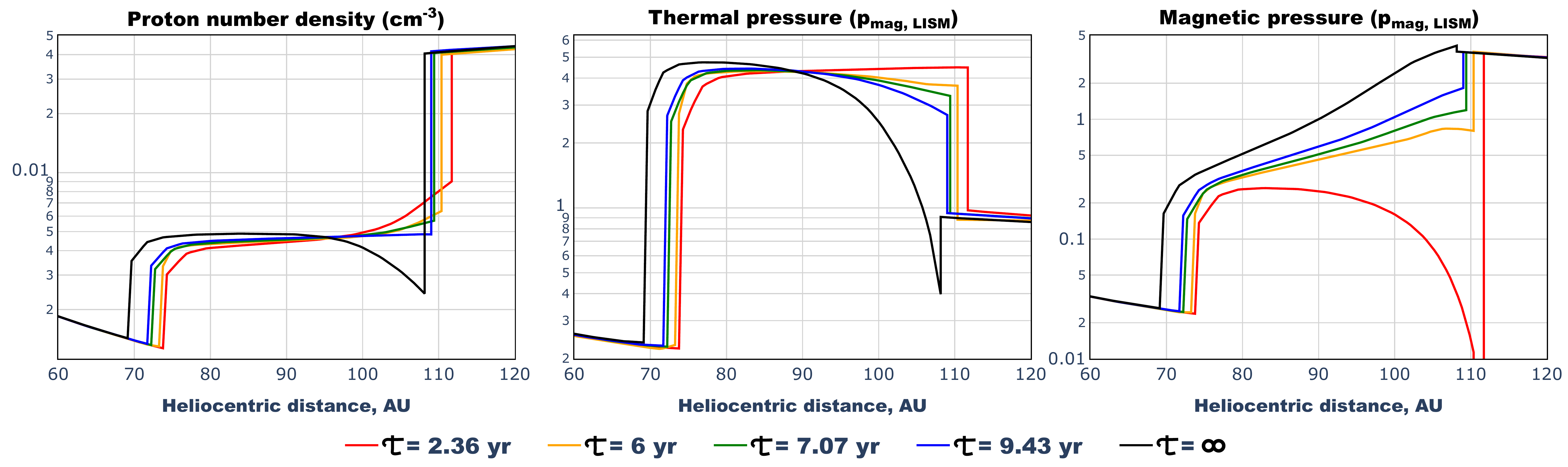}
\caption{
The model dependencies of the \change{proton} number density (left panel), thermal pressure (middle panel), and magnetic pressure (right panel) for a non-dissipative model ($\tau = \infty$, black lines) and models with dissipation timescales $\tau$ ranging from $\sim$2 to $\sim$10 years (colored lines) on the distance in the upwind direction. The thermal and magnetic pressures are normalized by the value p$_{\rm mag,LISM}$ = B$_{\rm LISM}^2 / (8\pi)$ of the magnetic pressure in the LISM.
}
\label{fig:parametric1}
\end{figure*}

In this section, we analyze the heliospheric plasma response to magnetic dissipation by varying the characteristic dissipation time $\tau$. Figure \ref{fig:parametric1} shows the model distribution along the upwind direction for three plasma parameters: proton number density (left panel), thermal pressure (middle panel), and magnetic pressure (right panel). These are shown for a non-dissipative model ($\tau = \infty$) and for models with $\tau$ from about 2 to 10 years. The figure demonstrates that increased magnetic dissipation efficiency in the IHS, which corresponds to a decrease in $\tau$, leads to three effects: (a) decreased magnetic pressure in the IHS, (b) increased thermal pressure near the HP, and (c) increased number density close to the HP. 

As discussed in the previous section, a timescale of approximately 6 years yields a reasonable agreement with the trends in magnetic pressure and proton number density in the IHS. This value was based on the following considerations. First, magnetic dissipation must be efficient enough to match the low magnetic pressure values observed by Voyager in the IHS. We determined that with $\tau$ > 6 years, dissipation is not strong enough to meet this criterion. Second, model results with $\tau$ < 6 years show pronounced magnetic pressure depletion near the HP and a gradual increase in proton number density at the HP (see red lines in the right and left panels of Figure \ref{fig:parametric1}). Both features are inconsistent with Voyager observations. 

\begin{figure}
\includegraphics[width=1.0\columnwidth]{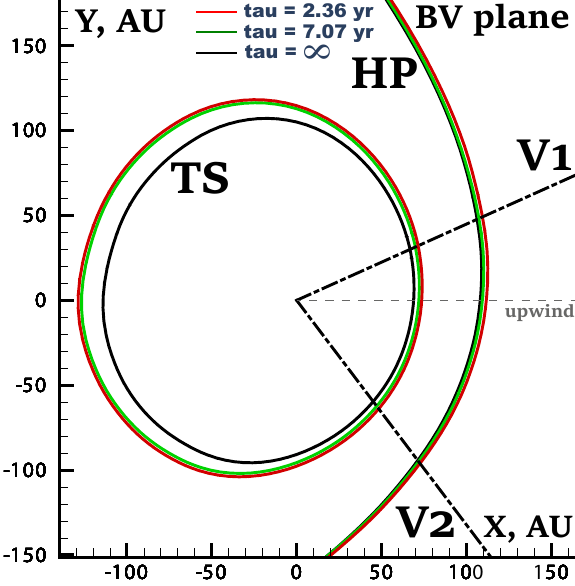}
\caption{
The heliopause (HP) and termination shock (TS) shapes in the BV-plane for models with different magnetic dissipation times: $\tau = \infty$ (no dissipation, black lines), $\tau = 7.07$ years (green lines), and $\tau = 2.36$ years (red lines). The directions towards the Voyager spacecraft (V1 and V2, dashed-dotted black lines) projected onto the BV-plane and the upwind direction (dashed black line) are shown for reference.
}
\label{fig:parametric2}
\end{figure}

Figure \ref{fig:parametric2} illustrates the positions of the TS and the HP in the BV-plane (the plane that contains $\mathbf{B}_{\rm LISM}$ and $\mathbf{V}_{\rm LISM}$ vectors) for models with $\tau = \infty$, 7.07, and 2.36 years. The latter two represent the cases of weak and strong magnetic field dissipation, respectively. The results indicate that magnetic dissipation increases the heliocentric distances of discontinuities (especially, the TS) and makes the IHS thinner. The most substantial change in heliocentric distances of TS is observed in the downwind region, where dissipation shifts the TS outward by $\sim$20 au. 

The HP position remains relatively stable at the flanks (particularly for $Y < 0$), while a more pronounced outward shift occurs in the upwind direction. This indicates that the change in the IHS width is direction-dependent, underscoring the inherently three-dimensional nature of the effect. The difference in the positions of discontinuities between $\tau = 2.36$ and $\tau = 7.07$ years model results is relatively small (compare red and green lines in Figure \ref{fig:parametric2}), suggesting that even weak magnetic dissipation provides a significant effect on the global structure of the heliosphere.

Direct comparison of the global heliosphere simulations with and without the HMF included was performed by \citet[][see their Models 1 and 2]{izmod2015}. These authors concluded that the effect of the magnetic wall and the plasma depletion around the HP in the model with the HMF included (which corresponds to $\tau = \infty$ in our notations) leads to a significant displacement of the TS toward the Sun (by $\sim$11 au in the upwind direction) compared to the model without the HMF (analogous to our $\tau \to 0$ case). This conclusion aligns well with our finding: the less effective the magnetic field dissipation, which corresponds to a higher value of $\tau$, the closer the TS to the Sun.

\section{Conclusions} \label{sec:conclusions}

Current global models of the heliosphere are unable to resolve magnetic field reconnection and the associated magnetic energy dissipation near the heliospheric current sheet due to its fine structure. In this study, we present an initial attempt to incorporate magnetic dissipation into a global model. We introduce a phenomenological dissipation term characterized by a dissipation time $\tau$, which operates within the inner heliosheath (IHS), into the magnetic induction equation. The proposed dissipation mechanism converts magnetic energy into the internal energy of the plasma, ensuring conservation of their total sum. We conducted a parametric analysis to examine the impact of the dissipation timescale $\tau$ on the global heliospheric structure. The principal findings are as follows:
\begin{enumerate}
    \item The mechanism of magnetic field dissipation suggested in this study (a) effectively reduces the magnetic pressure in the IHS and leads to a better agreement with V1 and V2 magnetometer observations, (b) eliminates the plasma density depletion near the HP caused by magnetic pile-up present in non-dissipative models of the heliosphere, resulting in a density trend consistent with V2 data, and (c) alters the bulk plasma flow insignificantly.

    \item The effect of magnetic field dissipation is responsible for the shift in the TS positions (outward from the Sun) and reduction in the width of the IHS. This is a step toward reconciling models with data, though it remains insufficient to resolve the problem of the IHS thickness.
    
    \item The characteristic magnetic dissipation time $\tau$ of approximately 6 years provides the best agreement with the trends of magnetic pressure and proton number density observed by the Voyager spacecraft in the inner heliosheath.
\end{enumerate}

Future studies will be devoted to the extension of this approach by (a) accounting for the dissipation efficiency dependence on the sector width (the separation between adjacent heliospheric current sheets along a streamline), and (b) accounting for not only dissipation but also the physical diffusion of the magnetic field, potentially informed by local, high-resolution models.

\section*{Acknowledgments}
\change{The work was supported by the research program <<Plasma>> of the Space Research Institute of Russian Academy of Sciences.}
The authors thank V. V. Izmodenov for fruitful discussions and D. B. Alexashov for providing the global distributions of plasma and hydrogen atoms in the heliosphere obtained using the \citet{izmod2020} model.
The research also benefited from discussions during the meeting of the International Space Science Institute (ISSI) team “Physical Processes and Drivers of Particle Acceleration in the Heliospheric Tail As Seen Through ENAs and Interstellar Lyman-alpha Absorption” (ISSI Team project №24-613, Lead: Kornbleuth M.).
The work of Merav Opher was supported by the NASA grant 18-DRIVE18\_2-0029 as part of the NASA/DRIVE program titled ``Our Heliospheric Shield,'' 80NSSC22M0164.
We acknowledge the NASA CDAWeb and OMNIWeb services (\url{https://cdaweb.gsfc.nasa.gov/}, \url{https://omniweb.gsfc.nasa.gov/}) for providing access to Voyager data and solar wind parameters at 1 au from the Sun, \change{and \citet{tokumaru2021} for making annual solar wind speed maps available at \url{https://stsw1.isee.nagoya-u.ac.jp/annual_map.html}.}

\bibliographystyle{jasr-model5-names}
\biboptions{authoryear}
\bibliography{bibliography}



\appendix

\section{Justification of the $\mathbf{Q}^B$ term} \label{app:Justification}

Using Faraday's and Ohm's laws (in cgs units)
\begin{equation}
    \nabla \times \mathbf{E} = -\frac{1}{c} \frac{\partial \mathbf{B}}{\partial t}, \: \mathbf{J} = \sigma (\mathbf{E} + \frac{1}{c} \mathbf{V} \times \mathbf{B}),
\end{equation}
where $\mathbf{J}$ is the current density, $\sigma$ is the electrical conductivity, the following equation can be written:
\begin{equation}
    \frac{\partial \mathbf{B}}{\partial t} = -c \nabla \times \mathbf{E} = \nabla \times (\mathbf{V} \times \mathbf{B}) - c \nabla \times \left( \mathbf{\frac{J}{\sigma}} \right).
\end{equation}
Since $\nabla \times (\mathbf{V} \times \mathbf{B}) = -\nabla \cdot (\mathbf{V} \mathbf{B} - \mathbf{B} \mathbf{V})$, we can derive the equation \ref{eq:induction}, where
\begin{equation}
    \mathbf{Q}^B = -c\nabla \times \left( \mathbf{\frac{J}{\sigma}} \right) = -\frac{c}{\sigma} \nabla \times\mathbf{J}, \label{eq:QB_general}
\end{equation}
and the last equality was obtained under the assumption that the electrical conductivity $\sigma$ is constant.

\begin{figure*}
\includegraphics[width=\textwidth]{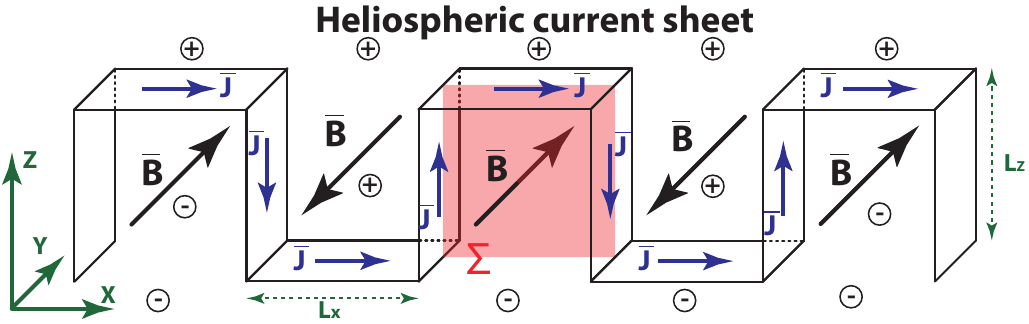}
\caption{
Schematic view of the heliospheric current sheet wavy structure in the inner heliosheath. The plus and minus signs denote the regions of the positive and negative heliospheric magnetic field polarity. The black arrows show the magnetic field direction, and the blue arrows represent the direction of the current density.
}
\label{fig:shem}
\end{figure*}

Let us consider the simplified geometry of the heliospheric current sheet shown in Figure \ref{fig:shem}. For convenience, we use the Cartesian coordinate system ($X$ and $Y$ are radial and tangential directions, respectively). We assume that (a) the magnetic field has only a tangential component ($\mathbf{B} = B_{\rm y} \mathbf{e}_{\rm y}$), and (b) $B_{\rm y}$ changes its sign at the HCS: above the current sheet the HMF polarity is positive ($B_{\rm y} < 0$), and below it is negative ($B_{\rm y} > 0$). 

To derive a simpler formula for the $\mathbf{Q}^B$ for the introduced planar HCS topology, we do the following steps.
\begin{enumerate}
    \item First, we find the value of the current density $\mathbf{J}$ in the HCS. According to Ampere’s law,
    \begin{equation}
        \mathbf{J} = \frac{c}{4\pi} \nabla \times \mathbf{B}.
    \end{equation}
    For segments of the HCS, which are parallel to the $XY$-plane, the current density has only $X$-component along the HCS: $J_{\rm x} \propto (\nabla \times \mathbf{B})_{\rm x} = -\partial B_{\rm y} / \partial z$. 
    We assume that the HCS has a thickness $w$. To calculate the magnitude of the current density $\mathbf{J}$, we approximate the corresponding derivative by the total change in $\mathbf{B}$ across the sheet and get the following formula:
    \begin{equation}
        |\mathbf{J}| \approx \frac{c}{4\pi} \frac{2 |\mathbf{B}|}{w}. \label{eq:J_module}
    \end{equation}
    The same formula could be obtained for other segments of the HCS, which are parallel to the $YZ$-plane.

    \item Second, we find $\nabla \times\mathbf{J}$. We apply Stokes' theorem
    \begin{equation}
        \int_{\Sigma} (\nabla \times \mathbf{J}) \cdot \mathrm{d}\mathbf{s} = \int_{\partial \Sigma} \mathbf{J} \cdot \mathrm{d}\mathbf{l}, 
    \end{equation}
    to the surface $\Sigma$ with normal $\mathbf{e}_{\rm y}$ (shown in red in Figure \ref{fig:shem}) and calculate the integrals as follows:  
    \begin{equation}
        \int_{\Sigma} (\nabla \times \mathbf{J}) \cdot \mathrm{d}\mathbf{s} = (\nabla \times \mathbf{J})_{\rm y} L_{\rm x} L_{\rm z},\: \int_{\partial \Sigma} \mathbf{J} \cdot \mathrm{d}\mathbf{l} = |\mathbf{J}| (2L_{\rm z} + L_{\rm x}),
    \end{equation}
    where $L_{\rm x}$ is the sector spacing in the $X$ (radial) direction (separation between two adjacent heliospheric current sheet folds) and $L_{\rm z}$ is the vertical size of the HCS. For the IHS, where the current sheet folding is happening, we have $L_{\rm x} \ll L_{\rm z}$, and, therefore, $(\nabla \times \mathbf{J})_{\rm y} \approx 2 |\mathbf{J}| / L_{\rm x}$. 
    
    If we apply Stokes' theorem to other surfaces within the sector zone with normals $\mathbf{e}_{\rm x}$ and $\mathbf{e}_{\rm z}$, we can find that $(\nabla \times \mathbf{J})_{\rm x} = (\nabla \times \mathbf{J})_{\rm z} = 0$. Therefore, assuming the simplified HCS geometry shown in Figure \ref{fig:shem}, the direction of $\nabla \times \mathbf{J}$ is the same as $\mathbf{B}$, and
    \begin{equation}
        \nabla \times \mathbf{J} \approx \frac{2 |\mathbf{J}|}{L_{\rm x}} \frac{\mathbf{B}}{|\mathbf{B}|}. \label{eq:rotJ_module}
    \end{equation}
\end{enumerate}

The combination of equations (\ref{eq:QB_general}), (\ref{eq:J_module}), and (\ref{eq:rotJ_module}) provides the dissipation term used in this work:
\begin{equation}
    \mathbf{Q}^B = -\frac{c}{\sigma} \nabla \times\mathbf{J} \approx -\frac{\mathbf{B}}{\tau},
\end{equation}
where the magnetic field dissipation timescale
\begin{equation}
    \tau = \frac{w L_{\rm x}}{4 \nu_{\rm m}},
\end{equation}
and $\nu_{\rm m} = c^2 / (4 \pi \sigma)$ is the magnetic viscosity. Therefore, the characteristic time for magnetic field dissipation $\tau$ is applicable only in the sector zone (the region of magnetic field polarity alternation), and should depend on the sector spacing $L_{\rm x}$. 

However, in the present study, $\tau$ is assumed constant throughout the IHS. \change{This assumption is adopted as a necessary simplification for our proof-of-concept investigation. A more physically accurate model would confine the dissipation to the compressed sector boundaries of the HCS, where reconnection is expected to dominate. Nevertheless, implementing such a localized dissipation presents a considerable challenge, as the HCS structure is inherently non‑stationary. Its highly warped and folded geometry varies substantially with solar cycle, precluding a simple, fixed definition in a steady‑state model. Developing a model that self‑consistently accounts for this dynamic morphology constitutes a natural extension of the present work.}

\section{Boundary conditions for the global model of the heliosphere} \label{app:boundary_conditions}

\subsection{Outer boundary conditions in the LISM}

The outer boundary conditions in the LISM are consistent with those of \citet{izmod2020}: the interstellar hydrogen, proton, and helium ions number densities are n$_{\rm H,LISM}$ = 0.14 cm$^{-3}$, n$_{\rm p,LISM}$ = 0.04 cm$^{-3}$, and n$_{\rm He+,LISM}$ = 0.003 cm$^{-3}$; the bulk velocity is V$_{\rm LISM}$ = 26.4 km s$^{-1}$, the direction of $\mathbf{V}_{\rm LISM}$ is longitude = 75°.4, latitude = -5°.2 in the ecliptic (J2000) coordinate system, the temperature T$_{\rm LISM}$ = 6530 K, the interstellar magnetic field magnitude is B$_{\rm LISM}$ = 3.75 $\mu$G, and the angle between $\mathbf{B}_{\rm LISM}$ and $-\mathbf{V}_{\rm LISM}$ equals 60° (so-called, BV-angle), where the interstellar magnetic field vector lies in the hydrogen deflection plane \citep[HDP,][]{lallement2005, lallement2010}. 

\subsection{Inner boundary conditions at 1 au from the Sun}

\begin{figure}
\includegraphics[width=\columnwidth]{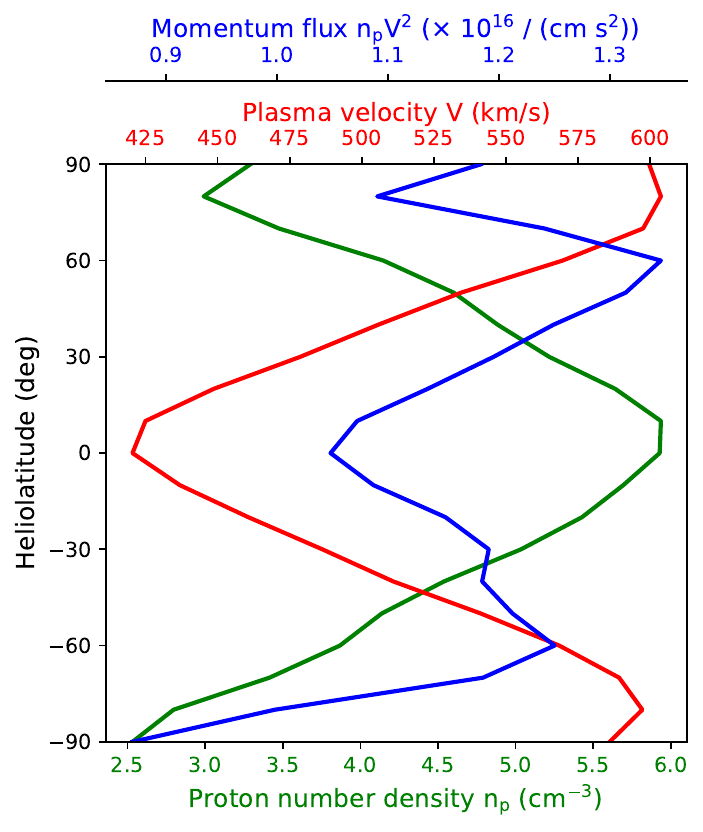}
\caption{
Solar wind proton number density (green curve), velocity (red curve), and momentum flux (blue curve) at 1 au as functions of the heliolatitude.
}
\label{fig:sw_profile}
\end{figure}

To obtain the inner boundary conditions for our stationary model, we follow the procedure described in Appendix A of \citet{izmod2020}. The heliolatitudinal variations of the SW proton density and velocity at 1 au were obtained using three different datasets:

\begin{enumerate}
    
    \item In the ecliptic plane, the SW proton density n$_{\rm p}$ and velocity $\rm V$ from the OMNI hourly averages dataset were used (\url{https://omniweb.gsfc.nasa.gov/}).

    \item Heliolatitudinal variations of the SW velocity $\rm V$ were taken from analysis of the interplanetary scintillation (IPS) data \citep{tokumaru2021}. Data are available yearly from 1985 to 2022 (except 2010; \url{https://stsw1.isee.nagoya-u.ac.jp/annual_map.html}).

    \item Heliolatitudinal variations of the SW mass flux ($\rm n_{\rm p} V$) were derived from analysis of SOHO/SWAN full-sky maps of the backscattered Lyman-$\alpha$ intensities \citep{katushkina2013, katushkina2019, koutroumpa2019}. An inversion procedure \citep{quemerais2006b} allows obtaining the SW mass flux as a function of time and heliolatitude with a temporal resolution of approximately 1 day and an angular resolution of 10$^\circ$. These data from 1996 to mid-2022 were provided by Dimitra Koutroumpa (personal communication).

\end{enumerate}

The heliolatitude profiles used in our stationary model simulations were obtained by averaging the time-dependent heliolatitudinal profiles over the period 2000.0 -- 2022.0. To determine the time-averaged heliolatitudinal velocity profile (red line in Figure \ref{fig:sw_profile}), the ratio of the time-averaged SW momentum flux ($\rm n_{\rm p} V^2$) to the averaged SW mass flux ($\rm n_{\rm p} V$) was calculated. The time-averaged number density profile (green line in Figure \ref{fig:sw_profile}) was calculated as the ratio of the time-averaged SW mass flux to the averaged velocity at a given heliolatitude.

At 1 au, we assumed that the number density of alpha particles (He$^{++}$) is 3.5194\% of the proton number density. The plasma temperature profile at 1 au was determined assuming that (a) the plasma thermal pressure is heliolatitudinally invariant, and (b) the Mach number at zero latitude is 6.408. For the heliospheric magnetic field, the unipolar Parker spiral solution has been assumed at 1 au with magnetic field radial component $B_{\rm R}$ = 37.5 $\mu$G \citep[for details, see][]{izmod2015}.

\section{Comparison of kinetic and multi-fluid descriptions of H atoms in global heliosphere modelling} \label{app:compare}

In this appendix, we present a comparison of plasma distribution from our non-dissipative model, which employs a multi-fluid treatment for H atoms, with the model by \citet{izmod2020}, which utilizes the conventionally accepted kinetic approach, under the same boundary conditions \citep[as in][]{izmod2020}. A similar comparative analysis for purely gas-dynamic models of the heliosphere (without the magnetic fields) was previously conducted by \citet{Alexashov2005} and \citet{heerikhuisen2006}.


\begin{figure*}
\includegraphics[width=\textwidth]{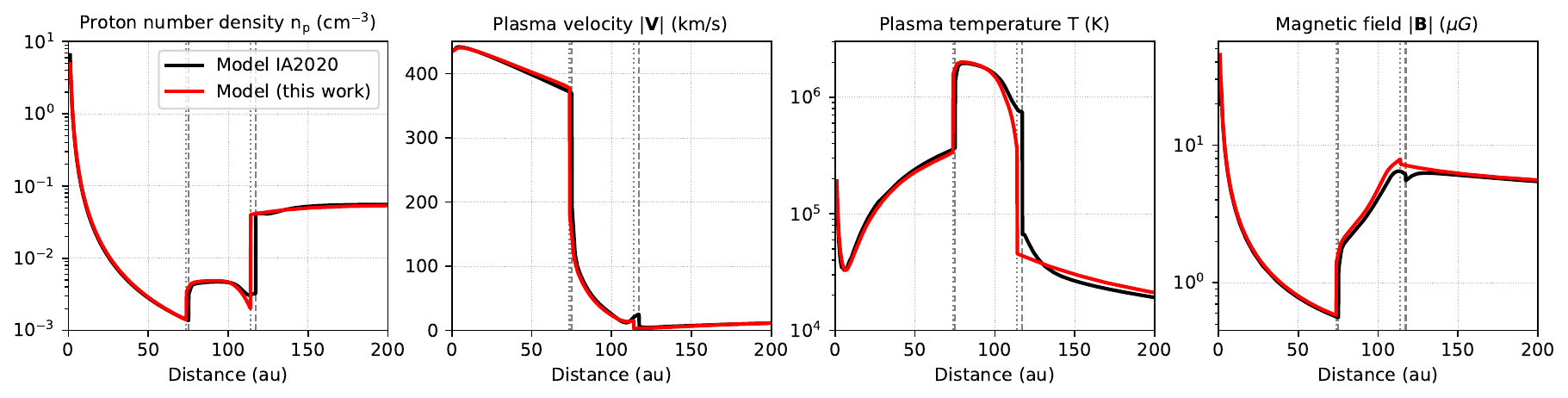}
\caption{
Distributions of the plasma parameters (proton number density, bulk velocity magnitude, plasma temperature, and magnetic field magnitude) along the upwind direction obtained using the \citet{izmod2020} model (black lines) and the non-dissipative model described in this work (red lines).
}
\label{fig:IA2020_comparison_1D}
\end{figure*}

Figure \ref{fig:IA2020_comparison_1D} compares the distributions of proton number density, bulk plasma velocity, plasma temperature, and magnetic field magnitude along the upwind direction for the two models. The black lines represent the results of \citet{izmod2020} model simulations, and the red lines represent the results of the non-dissipative model described in this work. The positions of the TS and HP differ by less than 3\% between the models in the upwind direction. As shown in this figure, there is a close agreement between the models in the upwind direction.

\begin{figure*}
\includegraphics[width=\textwidth]{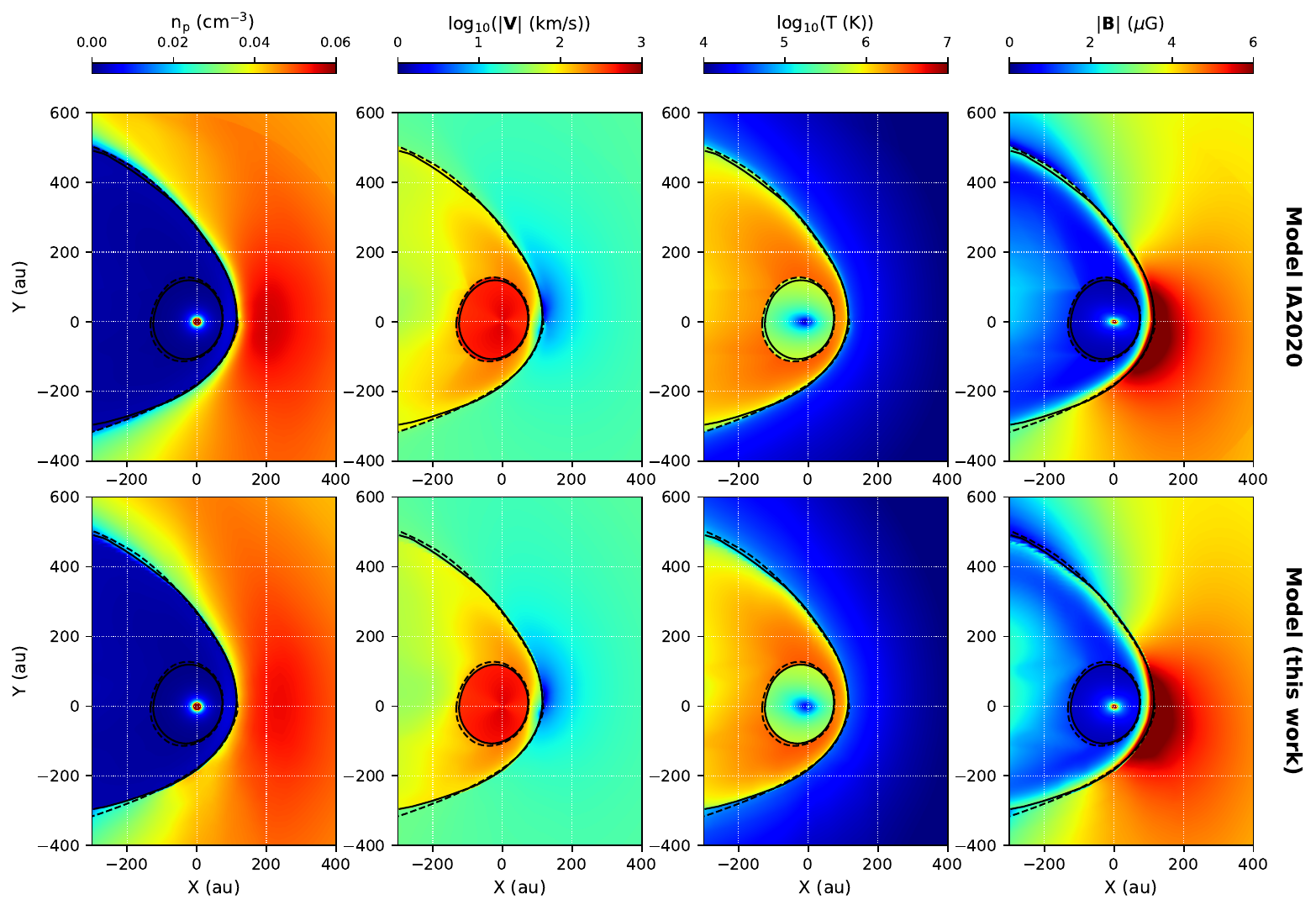}
\caption{
Distributions of the plasma parameters (proton density, bulk velocity magnitude, plasma temperature, and magnetic field magnitude) in the BV-plane obtained using the \citet{izmod2020} model (top row) and the non-dissipative model described in this work (bottom row). The TS and HP locations are shown with solid (model from this work) and dashed (\citet{izmod2020} model) black lines.
}
\label{fig:IA2020_comparison_2D}
\end{figure*}

Figure \ref{fig:IA2020_comparison_2D} presents 2D distributions of proton number density (first column), bulk plasma velocity (second column), plasma temperature (third column), and magnetic field magnitude (fourth column) for the two models in the BV-plane. The top row shows the results of the \citet{izmod2020} model simulations, and the bottom row represents the simulations of the non-dissipative model described in this work. The positions of the discontinuity surfaces (TS and HP) are overlaid, with solid lines representing the model from this work and dashed lines corresponding to the model by \citet{izmod2020}. As can be seen from this figure, the 2D comparison reveals nearly identical global structures and plasma distributions between the two models.

From the perspective of plasma dynamics, the multi-fluid model yields charge exchange source terms for momentum and energy that are broadly consistent (at least in an integral sense) with the kinetic model. Consequently, it reproduces a similar plasma flow. Although simpler to implement, it provides the locations of discontinuities and plasma parameter distributions that are close to those of the more physically comprehensive, but computationally expensive, kinetic approach.

While the multi-fluid model works well for plasma flow, it is important to note that the distributions of H atoms, particularly those originating in the solar wind region, can differ significantly from the kinetic results. This discrepancy may substantially impact the interpretation of observations, such as the Lyman-$\alpha$ absorption and emission spectra or energetic neutral atom (ENA) fluxes, as will be demonstrated in our future work. Given these differences in H atom distribution, we nevertheless focus our attention on the plasma properties in this work and conclude that the multi-fluid approach is adequate for the present investigation.

\end{document}